\documentclass[natbib]{svjour3}
\usepackage[dvips]{graphicx}
\newcommand{\eps}{\varepsilon}

\newcommand{\Kk}{{\cal K}}
\newcommand{\rb}{{\bf r}}

\newcommand{\Hd}{\overline{H}}
\newcommand{\mod}{\mathop{\rm mod}\nolimits}
\newcommand{\vphi}{\varphi}

\newcommand{\Rk}{{\cal R}}

\newcommand\beq[1]{ \begin{equation}\label{#1} }
\newcommand{\eeq}{ \end{equation} }

\newcommand\beqa[1]{ \begin{eqnarray}\label{#1}}
\newcommand{\eeqa}{ \end{eqnarray} }
\newcommand{\beqano}{ \begin{eqnarray*} }
\newcommand{\eeqano}{ \end{eqnarray*} }
\journalname{Celestial Mechanics and Dynamical Astronomy}
\title{Dynamics of ``jumping" Trojans: perturbative treatment}
\titlerunning{``Jumping" Trojans}
 \author{Vladislav V. Sidorenko}
 \institute{ V.V.Sidorenko
           \at Keldysh Institute of Applied Mathematics \\
               Russian Academy of Sciences, \\ \smallskip
               Miusskaya Sq., 4, 125047 Moscow, RUSSIA \\
               Moscow Institute of Physics and Technology \\
               Institutskiy S-Str., 9, 141700 Dolgoprudny, RUSSIA \\
               \email{vvsidorenko@list.ru}
 }
 \authorrunning{V.V.Sidorenko}
 \date{}
 
\begin{document}

 \bibliographystyle{plainnat}
 \maketitle
 \begin{abstract}
The term ``jumping'' Trojan was introduced by Tsiganis et al. (2000) in their studies of long-term dynamics exhibited by the asteroid (1868) Thersites, which had been observed to jump from librations around $L_4$ to librations around $L_5$. Another example of a ``jumping'' Trojan  was found by Connors et al. (2011): librations of the asteroid 2010 TK7  around the Earth's libration point $L_4$ preceded by its librations around $L_5$. We explore the dynamics of ``jumping'' Trojans under the scope of the restricted planar elliptical three-body problem. Via double numerical averaging we construct evolutionary equations, which allow analyzing transitions between the orbital motion regimes.
 \end{abstract}

 \section{Introduction}

An asteroid at 1:1 mean motion resonance with one of the main planets most often moves either in a ``tadpole'' orbit ($T$-orbit) or in a ``horseshoe'' orbit ($HS$-orbit). $T$-orbits cycle around one of the triangular libration points, whereas $HS$-orbits encompass both triangular libration points as well as the collinear libration point  $L_3$. Other types of resonance coorbital motion -- in particular, quasi-satellite ($QS$) regimes or compound $QS+HS$ orbits -- are also possible, although they are less common. The formal difference between these orbits is the behavior of the resonance phase $\vphi = \lambda - \lambda'$, where $\lambda$ and $\lambda'$ are the mean longitudes of the asteroid and the planet respectively \citep{Namounietal1999}.

If several modes of motion are possible for a Hamiltonian system at resonance, then under certain conditions the transitions between these modes can be observed. It was shown in \citep{Tsiganisetal2000} that Trojan asteroid (1868) Thersites will make a $T_L\rightarrow T_T$ transition ($T_L$ and $T_T$  denote $T$-orbits enclosing the ``leading'' and the ``trailing'' libration points $L_4$ and $L_5$ respectively). Numerical integration also indicates that the asteroid 2010TK7  (the first Trojan asteroid of the Earth) makes transitions between the motions in the neighborhood of  $L_4$  and  $L_5$ \citep{Connorsetal2011}. Further examples of similar ``jumps'' in the dynamics of real Trojans were discussed by \citet{FuenteMarcosandFuenteMarcos2012}, \citet{SchwarzandDvorak2012}, \citet{GalliazzoandSchwarz2014}.

Secular evolution of Trojan asteroids has been a point of interest for many specialists. A detailed bibliography can be found in
\citep{Erdi1997, Marzarietal2002, RobutelandSouchay2010}. The necessity to investigate the Trojans' jumps in simplified dynamical models was emphasized by \citet{SchwarzandDvorak2012}. For the time being, it seems that only \citet{OshimaandYanao2015} attempted an analytical study of the transitions
\beq{e0}
T_L \rightarrow T_T,\quad
T_T \rightarrow T_L,\quad
T_{L,T}\rightarrow HS,\quad
HS \rightarrow T_{L,T}.
\eeq
Their analysis was based mainly on the consideration of the planar restricted circular three-body problem. K.Oshima and T.Yahao ascribe the motions with transitions (\ref{e0}) to the region of chaotic dynamics generated by the intersection of stable and unstable manifolds of periodic solutions encircling the libration point $L_3$. However, the interpretation of transitions (\ref{e0}) as a certain homoclinic  phenomenon has a serious drawback -- the measure of the initial conditions giving rise to motions with transitions (\ref{e0}) turns out to be very small ($\sim \exp(-C/\sqrt{\mu})$, where $\mu$ characterizes the relative part of the planet's mass in the total mass of the system ``Sun+planet'',  $C=\mbox{const} > 0$; the presented estimate follows from some general results, obtained by \citet{Neishtadt1984}).

We aim to demonstrate that in the context of the planar restricted elliptic three-body problem ``Sun+planet+asteroid'' there is another mechanism underlying the transitions (\ref{e0}). To reveal this mechanism we apply the basic ideas of the approach proposed by J.Wisdom to study the transformations of the resonance motions \citep{Wisdom1985}. This approach also allows establishing the dynamical robustness of these transitions in the elliptic problem -- they occur for the set of initial conditions, whose measure does not depend on $\mu$. Previously and in a similar way we studied formation and destruction of $QS$ orbital motion regimes \citep{Sidorenkoetal2014}.

We hope our analysis to become a useful addition to the prior research on the secular effects in the dynamics of Trojan asteroids on the basis of the modern theory of resonance phenomena in Hamiltonian systems \citep{BeaugeandRoig2001, Morais2001}. Of course, the consideration of the three-body problem does not explain the transition  $T_L \rightarrow T_T$, demonstrated by the asteroid (1868) Thersites -- the numerical results presented by \citet{Tsiganisetal2000} indicate a significant influence of secular resonances on the dynamics of this asteroid. The mechanism of transitions that we are discussing is probably realized in the dynamics of the so called ``temporary'' Trojans \citep{Karlsson2004}. Since their stay in certain regimes of motion is relatively short, the effects due to secular resonances can be neglected.

\section{Averaged motion equations for studying the dynamics of the asteroid at 1:1 mean motion resonance}

\subsection{Averaging over orbital motion}

We assume the planet's orbit around the star to have a semimajor axis of unit length, and the sum of masses of the star and the planet to make the unit mass. The unit time is chosen so that the orbital period of the planet equals $2\pi$. The mass of the planet $\mu$ is substantially smaller than that of the star, and is further treated as a small parameter of the problem.

We focus our attention on the region ${\cal Z}_{res}$ of the system's phase space, defined by the condition
$$
\left|n-n' \right| \mbox{\raisebox{2pt}{$\mathop{<}\limits_{\displaystyle \sim}$}}\mu^{1/2}.
$$
Here $n$ and $n'=1$ are the mean motions of the asteroid and the planet respectively. The phase variables are
$$
x,\quad y,\quad L, \quad \varphi,
$$
where $x$, $y$, and $L$ are the Poincare elements, which are related to osculating elements by the formulae
\beq{e0a}
x=\sqrt{2\sqrt{(1-\mu)a}\left[1-\sqrt{(1-e^2)}\right]}\cos \varpi,
\eeq
$$
y=-\sqrt{2\sqrt{(1-\mu)a}\left[1-\sqrt{(1-e^2)}\right]}\sin \varpi,
$$
$$
L=\sqrt{(1-\mu)a}.
$$
Here $\varpi$, $e$, and $a$, are the longitude of the periapsis, the eccentricity, and the semimajor axis of the asteroid orbit respectively.

The equations of motion have the canonical form
\beq{e1}
\frac{dx}{dt}=-\frac{\partial {\cal K}}{\partial y},\quad
\frac{dy}{dt}=\frac{\partial {\cal K}}{\partial x},
\eeq
$$
\frac{dL}{dt}=-\frac{\partial {\cal K}}{\partial \vphi},\quad
\frac{d\vphi}{dt}=\frac{\partial {\cal K}}{\partial L},
$$
with the Hamiltonian
\beq{e2}
\Kk = - \frac{(1-\mu)^2}{2L^2}-L - \mu \Rk.
\eeq

The disturbing function $\Rk$ in the expression for $\Kk$ is defined as
$$
\Rk = \frac{1}{\left|\rb - \rb'\right|}- \frac{\left(\rb,\rb'\right)}{r'^3},
$$
where $\rb=\rb(x,y,L,\lambda(\vphi,\lambda'))$ and $\rb'=\rb'(\lambda')$ are the position vectors of the asteroid and the planet relative to the star.

Averaging of (\ref{e1}) over the orbital motion of the asteroid and the planet is equivalent to substituting the function
\beq{e3}
W(x,y,L,\vphi)=\frac{1}{2\pi}
\int_0^{2\pi} R(x,y,L,\lambda(\lambda',\vphi),\lambda')\,d\lambda'
\eeq
instead of the function $\Rk$ in the expression (\ref{e2}) for $\Kk$.

Such averaging eliminates the mean longitude of the planet $\lambda'=t+\lambda'_0$ from the right-hand sides of the equations of motion. Therefore, these equations become autonomous.

In our study the averaging (\ref{e3}) is carried out numerically. Technically, this is similar to the averaging of the disturbing function at 3:1 MMR described in detail in \citet{Sidorenko2006}. Let us note that numerical averaging of a disturbing function at MMR is a common technique (e.g., \citet{Schubart1964}).

\subsection{The ``slow-fast'' system}

We shall now proceed with the scale transformation
$$
\tau = \sqrt{\mu}t,\quad \Phi=(1-L)/\sqrt{\mu}.
$$
Without loss of accuracy the averaged equations of motion in the resonance zone ${\cal Z}_{res}$ can be rewritten as follows:
\beq{e4}
\frac{d\vphi}{d\tau}=3\Phi, \quad
\frac{d\Phi}{d\tau}=-\frac{\partial V}{\partial \vphi},
\eeq
$$
\frac{dx}{d\tau}=\eps \frac{\partial V}{\partial y}. \quad
\frac{dy}{d\tau}=-\eps \frac{\partial V}{\partial x}.
$$
Here
$$
\eps = \sqrt{\mu},\quad V(x,y,\vphi)=W(x,y,1,\vphi).
$$

Generally speaking, variables $x$, $y$, $\vphi$, and $\Phi$ in (\ref{e4}) vary with different rates:
$$
\frac{d\vphi}{d\tau},\frac{d\Phi}{d\tau}\sim 1, \quad
\frac{dx}{d\tau},\frac{dy}{d\tau}\sim \eps.
$$

Taking into account this separation of variables into fast and slow ones, we shall call the system (\ref{e4}) the ``slow-fast'' system (or SF-system). The ``fast'' subsystem consists of the equations for the variables $\vphi, \Phi$. The ``slow'' subsystem describes the behavior of the variables $x,y$.

SF-system (\ref{e4}) is a Hamiltonian one, whose symplectic structure is defined by the differential form
$$
\Psi=\eps^{-1}dy\wedge dx + d\Phi \wedge d\vphi.
$$
The corresponding Hamiltonian is
\beq{e5}
\Xi=\frac{3\Phi^2}{2}+V(x,y,\vphi).
\eeq

\subsection{Some relations between the function $V$ properties and the asteroid dynamics}

For $\eps=0$ the behavior of the ``fast'' variables is determined by 1DOF Hamiltonian system
\beq{e6}
\frac{d\vphi}{d\tau}=3\Phi, \quad
\frac{d\Phi}{d\tau}=-\frac{\partial V}{\partial \vphi},
\eeq
which depends on $x,y$ as parameters.

\begin{figure}[ht]
\vglue-7.0cm
\hglue4.0cm
\includegraphics[width=10cm,keepaspectratio]{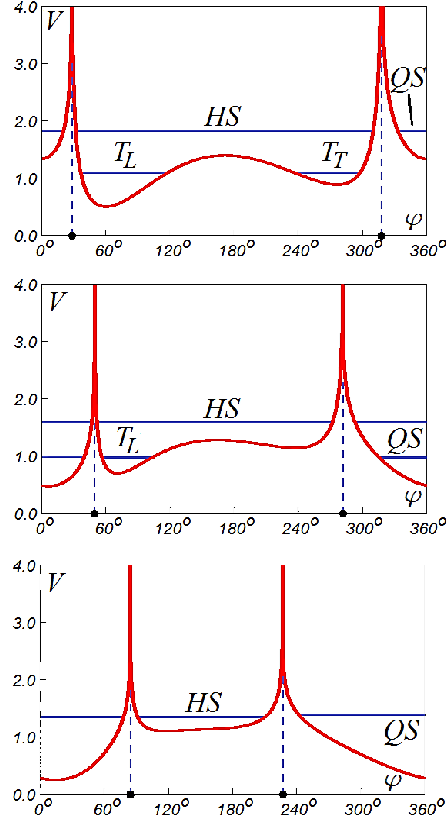}
\vglue0.0cm
\caption{Behavior of the function $V$ with fixed $x,y$. Upper panel: $x=0.15176$, $y=-0.26285$ ($e=0.3$, $\varpi=60^\circ$).
Middle panel: $x=0.31623$, $y=-0.54772$ ($e=0.6$, $\varpi=60^\circ$). Lower panel: $x=0.53100$, $y=-0.91987$ ($e=0.9$, $\varpi=60^\circ$). In all cases the eccentricity of the planet $e'=0.3$}
\label{p1}
\end{figure}

The properties of the solutions to the system (\ref{e6}) are determined by the function $V(x,y,\vphi)$ properties.
Figure \ref{p1} presents the graphs of this function for different values of $x,y$. The abbreviations $QS$,$HS$, and $T_{L,T}$ near the horizontal lines characterize the type of the secular evolution demonstrated by the asteroid in motions corresponding to solutions of (\ref{e6}) with a given value of the Hamiltonian $\Xi$ (in the limit case $\eps=0$). If
$x\neq \sqrt{2(1-\sqrt{1-{e'}^2})}$ and $y\neq 0$ (i.e., $e\neq e',\varpi \neq 0$), then in the interval $[0,2\pi]$ the function $V(x,y,\vphi)$ has two singular points $\vphi=\vphi_{SL}(x,y)$ and $\vphi=\vphi_{ST}(x,y)$ ($\vphi_{SL}(x,y)<\vphi_{ST}(x,y)$). If $x=\sqrt{2(1-\sqrt{1-{e'}^2})}$ and $y=0$ ($e=e',\varpi=0$), then $V(x,y,\vphi)\rightarrow +\infty$ at $\vphi \rightarrow 0 (\mod 2\pi)$. The singular points correspond to the motions of the asteroid ending up with its collision with the planet.

\begin{figure}[ht]
\vglue3.5cm
\hglue0.0cm
\includegraphics[width=10cm,keepaspectratio]{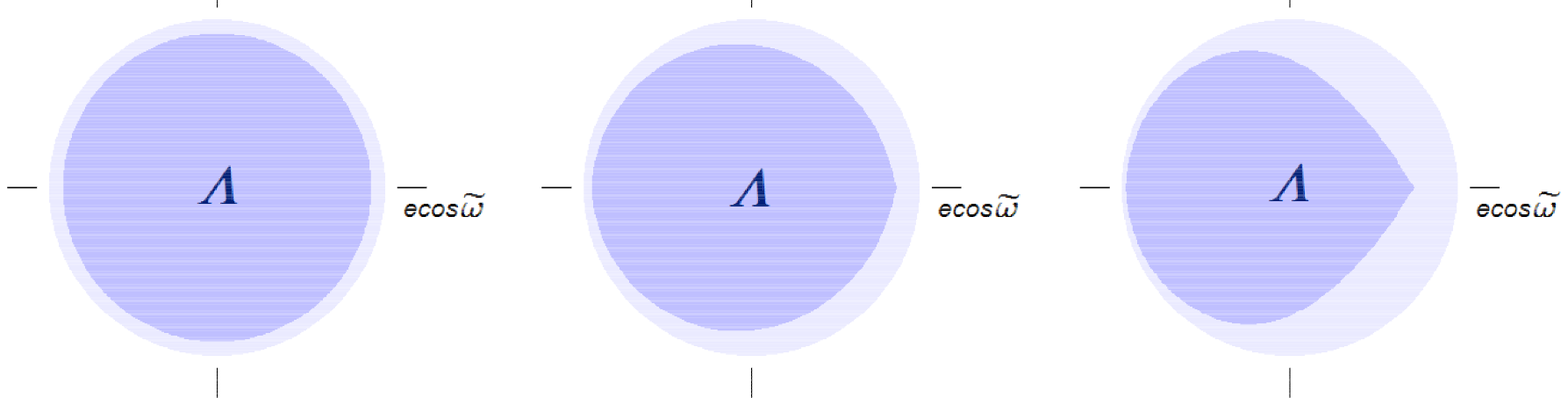}
\vglue-2.0cm
\caption{Set $\Lambda$ for different values of eccentricity of the planet. Left panel: $e'=0$. Middle panel: $e'=0.3$. Right panel: $e'=0.6$. We find it more clear to present diagrams and phase portraits in terms of the variables $e,\varpi$. To relate these variables with the variables $x,y$ we use the formulae (\ref{e0a}) with $\mu=0$, which is in accordance with the accuracy of our analysis}
\label{p2}
\end{figure}

Unlike the upper two graphs in Figure~\ref{p1}, the lower graph of $V(x,y,\vphi)$ does not have a bounded local maximum. Figure~\ref{p2} provides the examples of the set $\Lambda(e')$, consisting of the elements $x,y$, for which -- given the values of $e'$  -- the function $V(x,y,\vphi)$ has a bounded maximum (as a function of $\vphi$). The value of the resonance phase, which provides it, is denoted by $\vphi^*(x,y)$, assuming that $\vphi^*(x,y)\in(\vphi_{SL}(x,y),\vphi_{ST}(x,y))$. From the symmetry, inherent in the system, it follows that
\beq{e7}
V(v,y,\vphi)=V(x,-y,2\pi-\vphi)
\eeq
and consequently
$$
\vphi^*(x,0)=\pi.
$$

For $(x,y)\in \Lambda(e')$ the values of $\vphi$, for which $V(x,y,\vphi)$ has a minimum in the intervals $(\vphi_{SL}(x,y),\vphi^*(x,y))$ and $(\vphi^*(x,y),\vphi_{ST}(x,y))$  are denoted by $\vphi_{*L}(x,y)$ and $\vphi_{*T}(x,y)$ respectively.

If the eccentricities of the asteroid and the planet orbits are small, the following approximate formula can be applied:
\beq{e8}
V(x,y,\vphi) \approx V_0(\vphi)+V_1(x,y,\vphi).
\eeq
Here $\vphi \in [c_*,2\pi-c_*]$, $c_*$ is a positive constant, satisfying the condition $\max\{e,e'\}\ll c_* \ll 1$,
$$
V_0(\vphi)=\frac{1}{\sqrt{2(1-\cos\vphi)}}-\cos\vphi,
$$
$$
V_1(x,y,\vphi)=(x^2+y^2+{e'}^2)g_0(\vphi)+{e'}(xg_1(\vphi)+yg_2(\vphi)),
$$
$$
g_0(\vphi)=\frac{\cos\vphi}{2}+ \frac{9-5\cos^2\vphi - 4\cos\vphi}{4(2-2\cos\vphi)^{5/2}},
$$
$$
g_1(\vphi)=1-2\cos^2\vphi+\frac{\cos^3\vphi+8\cos^2\vphi-5\cos\vphi-4}{2(2-2\cos\vphi)^{5/2}},
$$
$$
g_2(\vphi)=2\cos\vphi\sin\vphi +\frac{\sin\vphi(9-\cos^2\vphi-8\cos\vphi)}{2(2-2\cos\vphi)^{5/2}}.
$$

The approximate expression (\ref{e8}) for the disturbing function is actually a special case of a more general formula obtained by \citet{Morais1999}, although the expression presented in \citep{Morais1999} lacks the term of the order ${e'}^2$.

\begin{figure}[ht]
\vglue-7.5cm
\hglue4.0cm
\includegraphics[width=10cm,keepaspectratio]{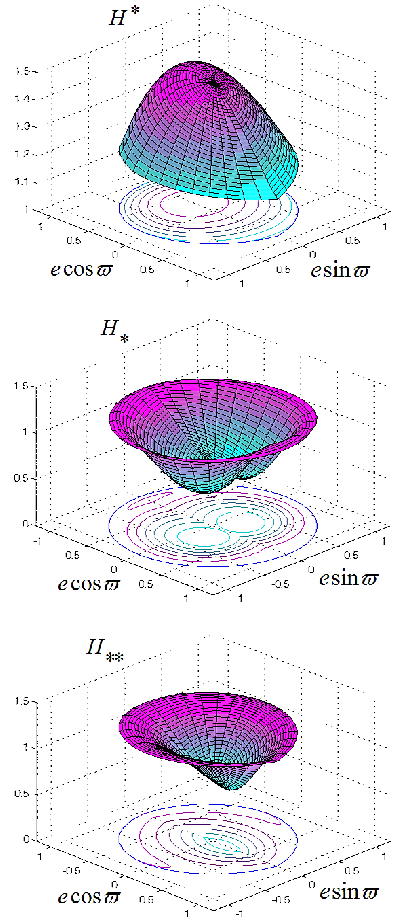}
\vglue0.5cm
\caption{Graphs of the functions $H^*,$ $H_*,$ and $H_{**}$ ($e'=0.3$)}
\label{p3}
\end{figure}

Following \citep{Sidorenko2006,Sidorenkoetal2014} we introduce the auxiliary functions
$$
H^*(x,y)=V(x,y,\vphi^*(x,y)),\quad H_*=\min_{\vphi\in(\vphi_{SL}(x,y),\vphi_{ST}(x,y))}V(x,y,\vphi),
$$
$$
H_{**}=\max\{V(x,y,\vphi_{*L}(x,y)),V(x,y,\vphi_{*T}(x,y))\}.
$$
The auxiliary functions $H^*(x,y)$ and $H_{**}$ are defined on $\Lambda(e')$, the function $H_*(x,y)$ is defined on the disk ${\cal D}=\{x^2+y^2<1\}$. Figure \ref{p3} presents sample graphs of the functions $H^*(x,y)$, $H_*(x,y)$, and $H_{**}(x,y)$ for $e'=0.3$.

If $e'\neq 0$, then the function $H^*(x,y)$ has a global maximum when
$$
x^*=-\sqrt{2(1-\sqrt{1-{e'}^2})},\; y^*=0.
$$
Taking into account the relations
$$
\frac{\partial H^*}{\partial x}=\left.\frac{\partial V}{\partial x}\right|_{\vphi=\vphi^*(x,y)},\;
\frac{\partial H^*}{\partial y}=\left.\frac{\partial V}{\partial y}\right|_{\vphi=\vphi^*(x,y)},\;
\left.\frac{\partial V}{\partial \vphi}\right|_{\vphi=\vphi^*(x,y)}=0,
$$
we conclude that SF-system (\ref{e4}) has a stationary solution
$$
x\equiv x^*,\; y\equiv 0,\; \vphi \equiv \pi,\; \Phi \equiv 0.
$$
This solution corresponds to the permanent stay of the asteroid at the collinear libration point $L_3$.

The function $H_*(x,y)$ has a minimum when
$$
x_*^{\pm}=\pm \sqrt{\frac{3(1-\sqrt{1-{e'}^2})}{2}},\;
y_* = \sqrt{\frac{3(1-\sqrt{1-{e'}^2})}{2}}.
$$
The corresponding stationary solutions to (\ref{e4})
$$
x\equiv x_*^{\pm},\; y\equiv y_*,\; \vphi \equiv \pm \frac{\pi}{3},\; \Phi \equiv 0,
$$
describe the stay of the asteroid at ``leading'' and ``trailing'' libration points respectively.

If the planet's eccentricity is sufficiently small, then the formula (\ref{e8}) allows obtaining an approximate expression for the function $H^*(x,y)$ in a neighborhood of the origin $(0,0)$:
\beq{e9}
H^*(x,y)\approx \xi^* + (x^2+y^2+{e'}^2)g_0(\pi)+e'x g_1(\pi)=\xi^*-\frac{7}{16}\left[(x+e')^2+y^2\right].
\eeq
Here
$$
\xi^*=V_0(\pi)=\frac{3}{2}.
$$
It is worth noting that $e' \ll 1$ entails
$$
x^*\approx -e'.
$$

\begin{figure}[ht]
\vglue2.5cm
\hglue0.0cm
\includegraphics[width=7cm,keepaspectratio]{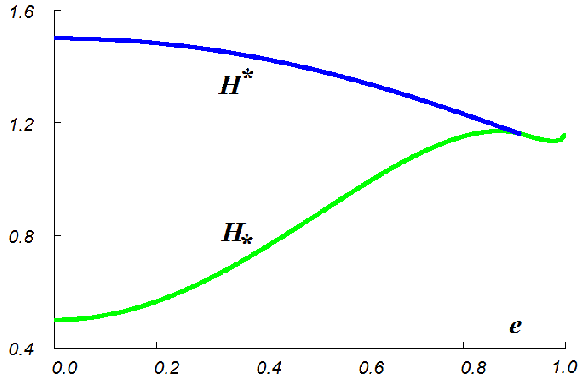}
\vglue-1.5cm
\caption{Graphs of the functions $H^*$ and $H_*$ in for $e'=0$}
\label{p4}
\end{figure}

For $e'=0$ the value of the auxiliary function at a point $(x,y)$ is determined by the distance from this point to the origin:
$$
H^*(x,y)=H^*(r),\; H_*(x,y)=H_*(r),\; H_{**}(x,y)=H_{**}(r),
$$
where $r=\sqrt{x^2+y^2}$. In the resonance zone ${\cal Z}_{res}$ with an accuracy of order $\eps$
$$
e=\sqrt{1-\left(1-\frac{r^2}{2}\right)^2}
$$
and, consequently, at $e'=0$ the auxiliary functions actually depend on the value of the asteroid's eccentricity $e$.
The graphs of the functions $H^*(e)$, $H_*(e)$ are shown in Figure \ref{p4}. The graph of the third auxiliary function is not given, since for $e'=0$
$$
H_{**}=H_*
$$
for all $(x,y)\in \Lambda(0)$.

\subsection{The solutions to the fast subsystem at $\eps=0$}

Let
\beq{e10}
\vphi(\tau,x,y,\xi),\; \Phi(\tau,x,y,\xi)
\eeq
denote a solution to equations (\ref{e6}), satisfying the condition
$$
\Xi(\vphi(\tau,x,y,\xi),\Phi(\tau,x,y,\xi),x,y)=\xi.
$$
In general, the angle $\vphi$ in the solution (\ref{e10}) oscillates with a period $T(x,y,\xi)$.

If $(x,y)\in \Lambda(e')$, then for $\xi \in (H_{**}(x,y),H^*(x,y))$ at the level set $\Xi=\xi$ there are two periodic solutions corresponding to motions in $T_L$-orbit ($\vphi_{SL}<\vphi<\vphi^*(x,y)$) and $T_T$-orbit ($\vphi^*<\vphi<\vphi_{ST}(x,y)$) respectively. In other cases a periodic solution, in which $\vphi \in (\vphi_{SL}(x,y),\vphi_{ST}(x,y))$, is either unique or does not exist at all.

We shall associate with the solution (\ref{e10}) the action integral
\beq{e11}
I(x,y,\xi)=\frac{3}{2\pi} \int_0^{T(x,y,\xi)}\Phi^2(\tau,x,y,\xi)\,d\tau.
\eeq

When $\eps \neq 0$ the variables $x(\tau),y(\tau)$ can be regarded as slowly varying parameters of the fast subsystem. Therefore, the relation (\ref{e11}) defines the adiabatic invariant (AI) of the SF-system (\ref{e4}).

\subsection{Averaging along the fast subsystem solutions}

Averaging along (\ref{e10}) the right-hand sides of the equations for the slow variables $x,y$ in the system (\ref{e4}) yields the evolutionary equations
\beq{e12}
\frac{dx}{d\tau}=\eps\left\langle \frac{\partial V}{\partial y} \right\rangle,\;
\frac{dy}{d\tau}=-\eps\left\langle \frac{\partial V}{\partial x} \right\rangle,
\eeq
where
\beq{e13}
\left\langle \frac{\partial V}{\partial \zeta} \right\rangle
=\frac{1}{T(x,y,\xi)}\int_0^{T(x,y,\xi)}\frac{\partial V}{\partial \zeta}
(x,y,\vphi(\tau,x,y,\xi))\,d\tau.
\eeq

Applying the averaging procedure (\ref{e13}) it is necessary to take into account that the solution (\ref{e10}), lying at the chosen level $\Xi=\xi$, can be aperiodic (in other words, it can correspond to the separatrix on the phase portrait of the fast subsystem). To distinguish these situations, we introduce the set
$$
\Gamma(\xi)=\left\{(x,y)\in \Lambda(\xi),\;H^*(x,y)=\xi\right\}.
$$
The set $\Gamma(\xi)$ consists of points with such coordinates $(x,y)$ that there exists an aperiodic solution (\ref{e10}) to (\ref{e6}). Similarly to \citep{Wisdom1985, Neishtadt1987a}, this set is further referred to as the uncertainty curve. A detailed discussion of the dynamical effects possible when the projection of the phase point
$$
{\bf z}(t)=(x(t),y(t),L(t),\vphi(t))^T
$$
onto the plane $x,y$ approaches the uncertainty curve $\Gamma(\xi)$ is given in \citep{Neishtadt1987b, NeishtadtandSidorenko2004, Sidorenkoetal2014}.

For simplicity we limit our analysis to the case when the uncertainty curve $\Gamma(\xi)$ is an oval. This is the case when $\xi \in (\xi^{**},\xi^*)$, where
$$
\xi^{**}=\max_{(x,y)\in \partial \Lambda} \Hd^*(x,y),
$$
$\Hd^*(x,y)$ denotes the continuous extension of the function $H^*(x,y)$ to the boundary of $\Lambda(e')$.
It follows from the formula (\ref{e9}), under the condition
$$
0 < \xi^* - \xi \ll 1
$$
and for small values of the planet eccentricity the uncertainty curve is close to a circle of radius
$$
R(\xi)=4\sqrt{\frac{\xi^*-\xi}{7}}
$$
with the center at the point $(-e',0)$.

Let $D(\xi)$ be a set of points lying inside the curve $\Gamma(\xi)$. If $(x,y)\in D(\xi)$, then there are two periodic solutions on the level $\Xi = \xi$, corresponding to the asteroid's motion in $T_L$- and $T_T$-orbits. Averaging (\ref{e4}) along these solutions yields, generally speaking, a different result. Thus, in the region $D(\xi)$ the evolutionary equations (\ref{e12}) have two families of phase trajectories describing the secular evolution of $T_L$-orbits and $T_T$-orbits respectively. Examples will be given in the next Section.

\section{Secular evolution and transitions between different types of the orbital motion}

Numerical investigation has shown that at $\xi=\xi_b \approx \xi^* - \frac{7}{4}$ a bifurcation occurs leading to a change in the number of fixed points of the system (\ref{e12}). For this reason we shall separately consider the cases
\beq{e14}
\xi \in (\xi^{**},\xi_b)
\eeq
and
\beq{e15}
\xi \in (\xi_b,\xi^*).
\eeq

\subsection{Secular evolution for $\xi \in (\xi^{**},\xi_b)$}

\begin{figure}[ht]
\hglue1.5cm
\includegraphics[width=6cm,keepaspectratio]{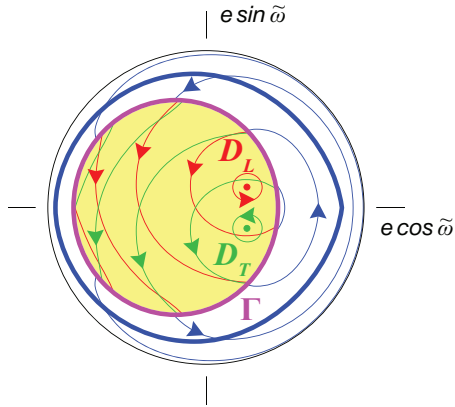}
\caption{Phase portrait of the system (\ref{e12}) for $\xi=1.3,$ $e'=0.3$}
\label{p5}
\end{figure}

Figure \ref{p5} presents a typical phase portrait of the system (\ref{e12}), when $\xi$ satisfies the condition (\ref{e14}). Trajectories approaching the uncertainty curve are formally glued with trajectories starting on this curve. At some points of the curve $\Gamma(\xi)$ two trajectories start simultaneously. It means that the projection of a phase point ${\bf z(t)}$ onto the plane $x,y$ can leave the neighborhood of $\Gamma(\xi)$ along any of these trajectories. This phenomenon allows a probabilistic interpretation (Section 4). A more detailed discussion of a similar situation can be found in \citep{Sidorenkoetal2014}, where temporal transitions to quasi-satellite orbits are considered under the scope of 3D RC3BP.

Red and green trajectories in Figure \ref{p5} characterize the secular evolution in the cases of the asteroid's motion in $T_L$-orbit and $T_T$-orbit respectively. Up to the arrows directions these families of the trajectories are symmetric with respect to the horizontal axis. For $(x,y) \in \Lambda(\xi')$ the blue trajectories correspond to the motion in $HS$-orbit, whereas outside of $\Lambda(e')$ the usual classification of orbits at $1:1$ MMR almost loses its sense.

\begin{figure}[ht]
\vglue4.2cm
\hglue.0cm
\includegraphics[width=10cm,keepaspectratio]{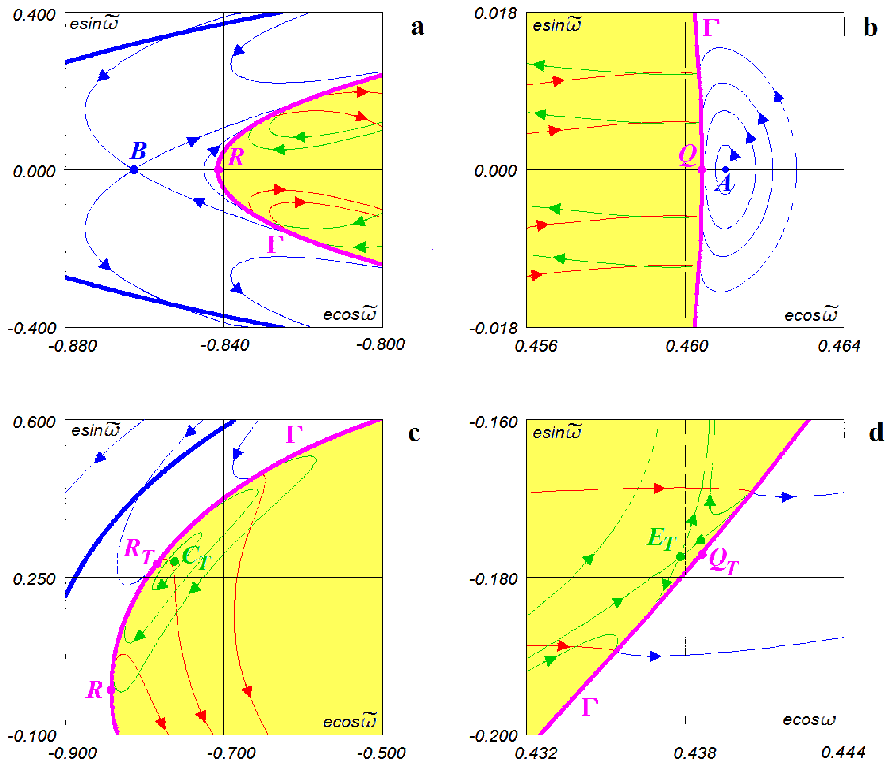}
\vglue-1.0cm
\caption{Behavior of the phase trajectories in the vicinity of the fixed points ($\xi=1.3,$ $e'=0.3$)}
\label{p6}
\end{figure}

There are eight fixed points on the described phase portrait. The fixed points $A$ and $B$ lie outside the region $D(\xi)$. The points $C_L$, $D_L$, and $E_L$ are the fixed points of the evolutionary equations (\ref{e12}) averaged along $T_L$-orbits. Similarly, the points $C_T$, $D_T$, and $E_T$ are the fixed points of (\ref{e12}) averaged along $T_T$-orbits. In Figure \ref{p5} only the fixed points $D_l$ and $D_T$ are shown, since the remaining points are very close to the curve $\Gamma(\xi)$. The enlarged fragments of the phase portrait in the vicinity of the points $A$, $B$, $C_T$, and $E_T$ are presented in Figure \ref{p6}. The behavior of the phase trajectories in the vicinity of the fixed points $C_L$ and $E_L$ is similar to the fragments with the points $C_T$ and $E_T$ respectively, up to the reflection with respect to the horizontal axis, followed by a change of colors and directions of the arrows.

Figure \ref{p6},a shows the behavior of phase trajectories in the vicinity of the unstable fixed point $B$ lying outside $D(\xi)$. The point $R \in \Gamma(\xi)$ is the limit point for a family of trajectories in the curvilinear triangle, whose sides are formed by the separatrices of the saddle point $B$ and the segment of the uncertainty curve $\Gamma(\xi)$.

Figure \ref{p6},b demonstrates the behavior of trajectories in the vicinity of the stable fixed point $A$ also lying outside the region $D(\xi)$. It can be seen, that as a limit, the family of trajectories, encircling this point, contains a trajectory tangent to the curve $\Gamma(xi)$ at the point $Q$.

Figure \ref{p6},c shows a vicinity of the fixed point $C_T$ of the averaged equations describing the evolution of $T_T$-orbits. The family of trajectories encircling $C_T$ includes as a limit a trajectory tangent to $\Gamma(\xi)$ at the point $R_T$.

Figure \ref{p6},d illustrates the behavior of the phase trajectories in the vicinity of the unstable fixed point $E_T$ of the equations (\ref{e12}) averaged along the $T_T$-orbits. The point $Q_T \in \Gamma(\xi)$ is the limit point for a family of trajectories in curvilinear triangle formed by the separatrices of the saddle point $E_T$ and the segment of the uncertainty curve $\Gamma(\xi)$.

The properties of the stable stationary solutions $A$, $C_L$, and $C_T$ have been given a closer look in \citet{Morais1999} for different values of a parameter that is actually equivalent to the parameter $\xi$.

\begin{figure}[ht]
\vglue1.0cm
\hglue3.0cm
\includegraphics[width=5.0cm,keepaspectratio]{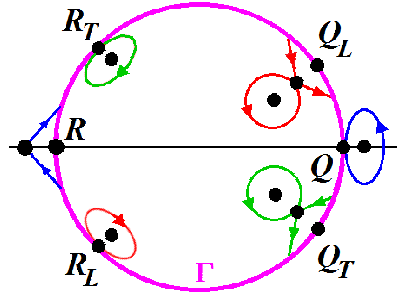}
\vglue-0.5cm
\caption{Partition of the curve $\Gamma(\xi)$ into segments with different motion transformations. The position of the points $Q$, $Q_L$, $Q_T$, $R$, $R_L$, $R_T$ is determined by the properties of some families of phase trajectories described in
Subsection 3.1}
\label{p7}
\end{figure}

\begin{figure}[ht]
\vglue3.5cm
\hglue0.0cm
\includegraphics[width=10cm,keepaspectratio]{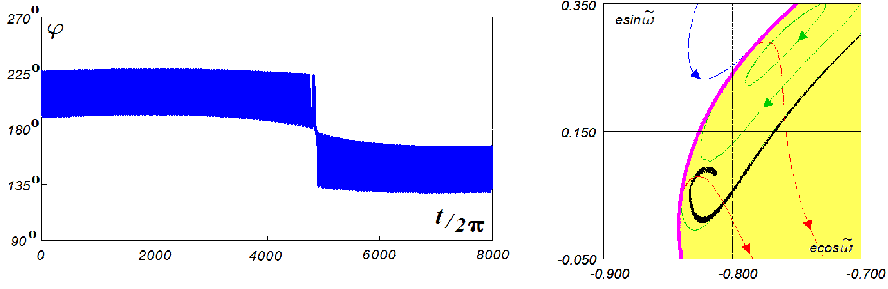}
\vglue-2.0cm
\caption{An example of a transition $T_T \rightarrow T_L$, obtained by numerical integration of non-averaged equations of motion ($\mu=0.0001,$ $e'=0.3$). The black curve on the right panel characterizes the evolution of the slow variables in this solution}
\label{p8}
\end{figure}

The points $Q$, $Q_L$, $Q_T$, $R$, $R_L$, $R_T$ generate the partition of the curve $\Gamma(\xi)$ into segments with different asteroid's motion transformations (Figure \ref{p7}). All possible scenarios are listed in the Table 1. As an example Figure \ref{p8} shows the results of numerical integration of non-averaged motion equations, demonstrating the transition $T_T \rightarrow T_L$ in the vicinity of the segment $R_T R$.

\noindent
\begin{center}
{\bf Table 1} Transformations of the motion regimes in the vicinity of the uncertainty curve
\vskip.1in

\begin{tabular}{ccccccc}
\hline
\makebox[1cm][c]{Segment} & $QQ_L$\rule[-7pt]{0pt}{22pt} & $Q_L R_T$ & $R_T R$ & $R R_L$ & $R_L Q_T$ & $Q_T Q$ \\
\hline
\makebox[1.1cm][c]{\parbox{2cm}{\begin{center}Possible transitions\end{center}}}\rule{0pt}{0pt}&
${\displaystyle HS \rule{0pt}{15pt}\atop \displaystyle T_L \rule{0pt}{15pt}}
{\searrow\rule[-5pt]{0pt}{7pt}\atop \nearrow} T_T$ &
$HS {\nearrow\rule[-5pt]{0pt}{7pt}\atop \searrow}
{\displaystyle T_L \rule{0pt}{15pt}\atop \displaystyle T_T \rule{0pt}{15pt}}$&
${\displaystyle HS \rule{0pt}{15pt}\atop \displaystyle T_T \rule{0pt}{15pt}}
{\searrow\rule[-5pt]{0pt}{7pt}\atop \nearrow} T_L$&
$T_T {\nearrow\rule[-5pt]{0pt}{7pt}\atop \searrow}
{\displaystyle HS \rule{0pt}{15pt}\atop \displaystyle T_L \rule{0pt}{15pt}}$&
${\displaystyle T_L \rule{0pt}{15pt}\atop \displaystyle T_T \rule{0pt}{15pt}}
{\searrow\rule[-5pt]{0pt}{7pt}\atop \nearrow} HS$&
$T_L {\nearrow\rule[-5pt]{0pt}{7pt}\atop \searrow}
{\displaystyle HS \rule{0pt}{15pt}\atop \displaystyle T_T \rule{0pt}{15pt}}$
\\
\hline
\end{tabular}
\vskip.1in
\end{center}

{\it Note.} Numerical studies did not reveal any other motion transformations in the vicinity of $\Gamma(\xi)$ for $\xi \in (\xi^{**},\xi^*)$. The value of $\xi$ affects only the position of points $Q$, $Q_L$, $Q_T$, $R$, $R_L$, $R_T$ on the uncertainty curve and the position of the uncertainty curve in the plane of the slow variables.

\vskip.1in

If the uncertainty curve encircles the origin $(0,0)$ in the plane of the slow variables, then the longitude of periapsis $\varpi$ can be used as a parameter that determines the position of the phase point on $\Gamma(\xi)$:
$$
\varpi = \left\{
\begin{array}{cr}
2\pi - \arccos\frac{x}{\sqrt{x^2+y^2}},& y\geq 0; \\
  \arccos\frac{x}{\sqrt{x^2+y^2}}\rule{0pt}{17pt},& y < 0. \\
\end{array}
\right.
$$

\begin{figure}[ht]
\vglue2.0cm
\hglue2.0cm
\includegraphics[width=6.0cm,keepaspectratio]{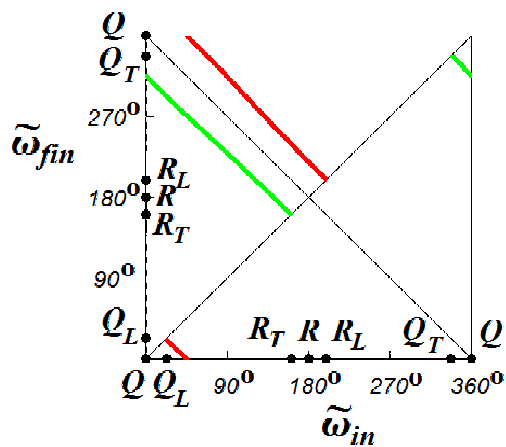}
\vglue-1.0cm
\caption{A correspondence between the value of the longitude of periapsis just after the transition and its value just before the next transition for motion in $T$-orbit ($\xi=1.3,$ $e'=0.3$). Red and green curve correspond to $T_L$-regime and $T_T$-regime respectively}
\label{p9}
\end{figure}

The phase flow of the averaged equations generates a map $\Gamma \rightarrow \Gamma$. For motion in $T$-orbit it is easy to establish a correspondence between the initial value of $\varpi$ (i.e., just after the transition to this orbit) and its value, when the transition to another orbital regime takes place. An example is presented in Figure \ref{p9}. It is noteworthy that this correspondence is not uniquely defined: for some ``input'' values of the longitude of periapsis $\varpi_{in}$ the next approach to the curve $\Gamma(\xi)$ is possible with two different values $\varpi_{fin}$.

\subsection{Secular evolution for $\xi \in (\xi_b,\xi^*)$}

\begin{figure}[ht]
\hglue1.5cm
\includegraphics[width=6cm,keepaspectratio]{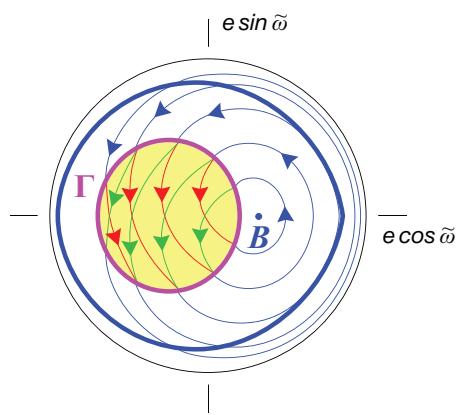}
\caption{Phase portrait of the system (\ref{e12}) at $\xi=1.4,$ $e'=0.3$}
\label{p10}
\end{figure}

A typical phase portrait of the system (\ref{e12}) for $\xi \in (\xi_b,\xi^*)$ is shown in Figure \ref{p10}. Its main difference from the phase portrait in Figure \ref{p5} is the absence of the fixed points $D_{L,T}$ and $E_{L,T}$. Nevertheless, the pairwise merging of these points at $\xi=\xi_b$ does not affect the behavior of the phase trajectories in the vicinity of the uncertainty curve. And, consequently, there are no qualitative changes in the partition of $\Gamma(\xi)$ into segments with different motion transformations.

\section{The transitions between different types of the co-orbital motion: probabilistic characteristics}

\begin{figure}[ht]
\vglue3.5cm
\hglue0.0cm
\includegraphics[width=10cm,keepaspectratio]{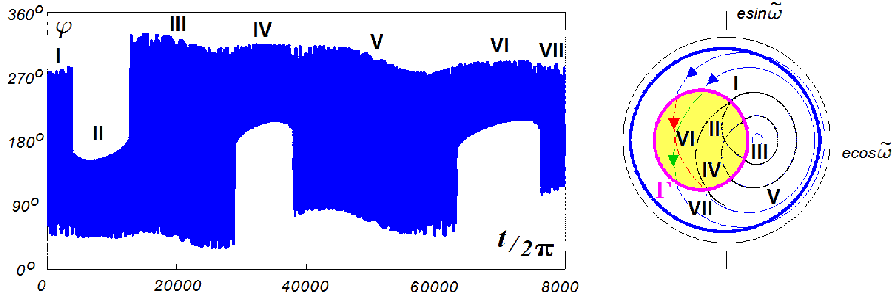}
\vglue-2.0cm
\caption{Transitions between different types of motion on the uncertainty curve $\Gamma$. Black curve on the right panel characterizes the evolution of the slow variables in this solution of the non-averaged equations ($\mu=0.00002,$ $\xi=1.4,$ $e'=0.3$). Segments $I$ and $V$ of the presented solution practically coincide (up to a quasi-random component of AI variation in the vicinity of $\Gamma$) and end at close points of the uncertainty curve}
\label{p11}
\end{figure}

Figure \ref{p11} presents the results of numerical integration of non-averaged motion equations demonstrating the transition to different orbital regimes when the projection of the phase point ${\bf z}(t)$ onto the plane of the slow variables intersects the uncertainty curve virtually in the same place. The fact, that the qualitatively different variants of the secular evolution can be realized, when the phase points leave the vicinity of $\Gamma(\xi)$, means faster ``chaotization'' of the dynamics of the system in comparison with the case when the ``scattering'' of the trajectories in the vicinity of $\Gamma(\xi)$ is associated only with a violation of adiabaticity (as, for example, in \citep{Neishtadt1987a,Neishtadt1987b,Wisdom1985}).

In case of a strong mixing of the initial conditions corresponding to different dynamical regimes, the probabilistic estimates of the possible motion transformations become meaningful. Indeed, even a small uncertainty in the initial conditions does not allow to predict uniquely the qualitative character of the motion over long time intervals. Not being rigorous enough, we define the probability of a certain motion regime as the relative measure of the set of initial conditions leading to this regime in the sufficiently small region of the phase space. Strict definition is given in \citep{Arnold1963, Neishtadt1987b}.

To obtain the transition probabilities we calculate the values of the auxiliary quantities introduced in \citep{Neishtadt1987b, Artemyevetal2013}:
$$
\Theta_{L,T}=\int_{-\infty}^{+\infty}
\left(
\frac{\partial H^*}{\partial x}\frac{\partial V}{\partial y} -
\frac{\partial H^*}{\partial y}\frac{\partial V}{\partial x}
\right)_{\vphi^s_{L,T}(\tau,x,y)}
d\tau.
$$
Here $\vphi^s_{L,T}(\tau,x,y)$ denotes aperiodic solutions of the fast subsystem lying on the critical level $\Xi=H^*(x,y)$.

The quantities $\Theta_L$ and $\Theta_T$ are time derivatives of the area of the regions bounded by the separatrices on the phase portraits of the fast subsystem, when the slow variables evolve according to the equations (\ref{e12}).

If the projection of the phase point ${\bf z}(t)$ onto the plane of slow variables approaches the uncertainty curve $\Gamma(\xi)$, then the probabilities of the subsequent motion regimes are given by formulae by \citet{Artemyevetal2013}:
\beq{e16}
P_{L,T}=\frac{\hat{\Theta}_{L,T}}{\hat{\Theta}_L+\hat{\Theta}_T+\hat{\Theta}},\quad
P_{HS}=1- P_L - P_T,
\eeq
where $\hat{\Theta}_{L,T}=\max(\Theta_{L.T},0)$, $\hat{\Theta}=\max(-\Theta_L-\Theta_T,0)$.

\begin{figure}[ht]
\vglue2.5cm
\hglue1.0cm
\includegraphics[width=8cm,keepaspectratio]{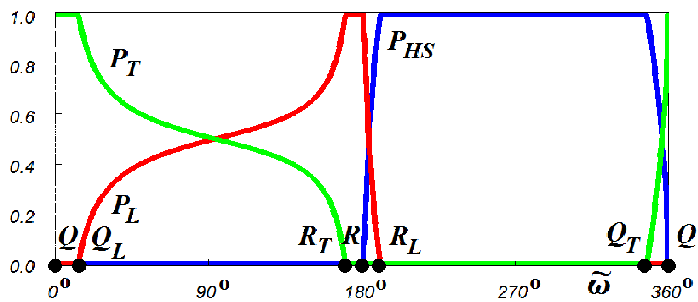}
\vglue-1.5cm
\caption{Probabilities of different types of co-orbital motion after approaching the curve $\Gamma$ ($\xi=1.3,$ $e'=0.3$)}
\label{p12}
\end{figure}

Figure \ref{p12} provides an example of the typical probability distribution of various transitions along the curve $\Gamma(\xi)$. Calculations were carried out using the formulae (\ref{e16}) for the case $\xi=1.3$, $e'=0.3$.

\section{Conclusion}

We presented a detailed investigation of the possible motion transformations at 1:1 MMR in the context of the planar restricted elliptic three-body problem. Similar transformations of the resonance regimes occur at other MMR as well. Examples can be found in \citep{ChiangandJordan2002, Ketchumetal2013}. To describe the motion with transitions between different resonance regimes \citet{Ketchumetal2013} proposed a rather illustrative term ``nodding behavior''. We consider the analysis of ``nodding behavior'' as a very interesting trend in the studies on MMR and hope to have contributed to its progress.

\section*{Acknowledgments}

The work was supported by the Presidium of the Russian Academy of Sciences (Program 7 ``Experimental and theoretical studies of the objects in the Solar system and exoplanetary systems''). We are grateful to S.S.Efimov, B.Erdi, A.I.Neishtadt, and D.A.Pritykin for reading the manuscript and useful discussions.

\end{document}